\documentclass[11pt,a4paper]{article}
\usepackage{jheppub_kim}
 \usepackage{slashed}

\usepackage{longtable}

\usepackage{epsfig}
\usepackage{amsmath}
\usepackage{graphicx}
 \usepackage{graphics}
 \usepackage[scriptsize,nooneline,hang]{caption}
\usepackage[hang,nooneline,scriptsize]{subfigure}

\usepackage{array}
\usepackage{makecell}

\begin{document}
\title{\Large   Soft-Collinear Effective  Theory: BRST Formulation}

 \author[a,b]{ Sudhaker Upadhyay }
 \author[c]{and Bhabani Prasad Mandal}
\affiliation[a]{Department of Physics, K.L.S. College,  Nawada-805110, (a constituent unit of Magadh University, Bodh-Gaya), Bihar, India}
\affiliation[b]{Visiting Associate, Inter-University Centre for Astronomy
and Astrophysics (IUCAA) Pune-411007, Maharashtra, India}
\affiliation[c]{Department of Physics, Institute of Science, Banaras Hindu University, Varanasi-221005 India}
\emailAdd{sudhakerupadhyay@gmail.com; sudhaker@associates.iucaa.in}
 \emailAdd{bhabani.mandal@gmail.com}
 
\abstract{We provide a systematic BRST formalism for the soft-collinear effective theory 
describing   interactions of soft and collinear degrees of freedom in the presence of a hard interaction. In particular, we develop full BRST symmetry transformation for SCET theory.  We further extend the BRST formulation by making the transformation field dependent. 
This establishes  a mapping  between several
SCET actions   consistently when defined in different
gauge conditions.  In fact, a definite structure of gauge-fixed actions corresponding to
any particular gauge condition can be generated for SCET theory using our formulation.}
\keywords{ Soft-Collinear Effective  Theory; BRST symmetry; Extended BRST symmetry.}
\maketitle
\section{Introduction}
 Effective field theories  are used to separate the the contributions
associated with different scales, a high-energy    and a  low-energy scales, of Quantum Chromodynamics (QCD).
  Over the past two decades, soft-collinear effective field theory (SCET)
\cite{1,2,22} has become amongst one of the important theories describing
 low-energy effective field theories of the Standard Model. 
 In  QCD, the low-energy part is  nonperturbative in particular.
In order to derive
the factorization theorems and to perform the resummation of Sudakov logarithms, SCET provides an alternative to the traditional diagrammatic techniques \cite{3}.  
SCET  has been applied to a large variety of
processes, from $B$-meson decays to jet production at the Large Hadron Collider (LHC).
In Ref. \cite{22}, the factorization of soft and ultrasoft gluons from collinear particles is
shown at the level of operators.
 
In order to describe jet-like events of QCD in SCET, it is convenient to 
write fields in either collinear, anti-collinear or soft (low energetic) modes
with the help of the light cone unit vectors 
satisfying $n^2=\bar n^2=0$ and $n\cdot\bar n=2$. A momentum  in the light-cone basis 
is represented as
\begin{eqnarray}
p^\mu &=&\frac{n^\mu }{2}\bar n\cdot p +\frac{\bar n^\mu }{2}  n\cdot p + p_\perp^\mu,
\nonumber\\
&=& \frac{n^\mu }{2}p^- +\frac{\bar n^\mu }{2}  p^+ + p_\perp^\mu,
\end{eqnarray}
  where $\perp$  components are orthogonal to both collinear unit vector $n$ and anti-collinear unit vector $\bar n$.
  
The  gauge symmetry structure in SCET
is richer than the QCD as the former  involves more than one distinct gluon fields.
 Therefore, the idea of background fields is required
to give well defined meaning to several distinct gluon fields \cite{4}.
Based on momentum regions, SCET is categorized in two formulations: SCET I and SCET II. 
SCET I and SCET II scale soft sector of the theory differently. For instance, in SCET I  all the momentum components of the soft fields
 are scaled  similar to the small component of the collinear
fields, while in SCET-II the momentum components of   soft fields are scaled like the
transverse component of the collinear fields.  

The celebrated  Becchi--Rouet--Stora--Tyutin (BRST) formulation   is  a comparatively rigorous mathematical  scheme \cite{brs1,brs2,brs3} which provides  a powerful technique to quantise gauge field theories. The range of applicability 
of BRST formulation further enhanced by extending it, where the anti-commuting  transformation parameter is made  
 finite and field-dependent  \cite{sdj}.
The finite field-dependent BRST transformations have been discussed
successfully in  many  field theoretic systems with gauge symmetries and have been found many applications  \cite{sdj1,ssm,sm,s,sm1,sm2,s1,s2,s3,s4,s5,s6,s7,s8,s9}.
Although
 BRST formulation simplifies the   renormalizability greatly
and helps to show unitarity of many  theories, the implementation of this approach in SCET is quite cumbersome task. Thus even though  a full field theoretic description for 
hadronic processes is developed,  the BRST formulation for SCET is not studied so far. 
This provides us with an opportunity to bridge this gap.

In this paper, we consider a gauge invariant SCET I action which admits different sets of
 gauge invariance in different momentum regions. We develop two sets of BRST symmetries which leave the Faddeev-Popov actions for collinear and ultrasoft sectors, separately. Moreover,
 we formulate an extended version of BRST symmetries by making the transformation parameter
 field dependent. We call such transformation field-dependent BRST (FDBRST) transformation.
  In contrast to the standard case, this eventually leads to a non-trivial  
 Jacobian for functional measure in the expression of transition amplitude. This Jacobian
 extends the BRST-exact parts of the action. We show that
 for some appropriate choices of field-dependent parameters an exact form of gauge-fixed action 
 corresponding to different gauge condition can be generated through FDBRST.

The plan of the paper is as following. In section \ref{2}, we construct fermionic rigid collinear and 
ultrasoft BRST transformations. These symmetry transformations are further generalized by making he transformation parameters field dependent in a traditional way in section \ref{3}. Moreover, we
implement such FDBRST transformations with appropriately constructed transformation parameters
to the generating functional. 
We summarize the outcome of this formulation with their significance in the section \ref{4}.

\section{SCET I action and BRST symmetry}\label{2}

In the  Ref. \cite{scet}, it has been shown that  the leading-order SCET collinear quark action  should  satisfy following requirement:
(a) it should yield   proper spin structure of the collinear propagator, (b)
it should have both collinear quarks and collinear antiquarks, (c) it should
  interact   with both collinear gluons and ultrasoft gluons, (d) and  it should
lead to the correct low-order propagator for different situations.  
These requirements allow  us to write down the effective leading-order SCET action. 
Further by splitting the fermion field into big and small components
using the usual projectors ($\frac{\slashed{n}\slashed{\bar n} }{4}$ and $\frac{\slashed{\bar 
n}\slashed{ n} }{4}$) and
eliminating (using the equations of
motion) the small components, one can write first the leading-order collinear quark action with collinear modes in   $n$ direction as \cite{scet}
\begin{eqnarray}
S_{n\xi}=\int d^4x \left[e^{x\cdot \mathcal{P}}\bar\xi_n\left(in\cdot D +i {\slashed{D}}
_{n\perp}\frac{1}{i\bar n\cdot D_n}i \slashed{D}_{n\perp}\right)\frac{\slashed{\bar n}}{2}
\xi_n\right],\label{fer}
\end{eqnarray}
where   $\mathcal{P}^\mu$ is a label operator which provides a definite power counting for derivatives and the collinear covariant derivatives are defined as
\begin{eqnarray}
i n\cdot D&=&in\cdot\partial +gn\cdot A_n+gn\cdot A_{us},\nonumber\\
i \slashed{D}^\mu_{n\perp}&=&\mathcal{P}_\perp^\mu+g A^\mu_{n\perp},\nonumber\\
i\bar n \cdot D_n&=&\bar{\mathcal{P}}+g\bar{n}\cdot A_n.
\end{eqnarray}
Even in the presence of ultrasoft fields, one can
write collinear quark action equivalent to (\ref{fer}) as
\begin{eqnarray}
S_{n\xi}=\int d^4x \left[e^{x\cdot \mathcal{P}}\bar\Xi_n i\slashed{\mathcal{D}}
\Xi_n\right],\  \ \Xi_n\equiv \begin{pmatrix}
\xi_n\\ \varphi_{\bar{n}}
\end{pmatrix}
\end{eqnarray}
where spinor components  $\varphi_{\bar{n}}$  are subleading in the collinear limit and
\begin{eqnarray}
 i\slashed{\mathcal{D}}=\frac{\slashed{\bar n}}{2} in\cdot D+\frac{\slashed{ n}}{2} i\bar n
 \cdot D_n+i \slashed{D}
_{n\perp} =i\slashed{D}_n+\frac{\slashed{\bar n}}{2}gn\cdot A_{us}.
\end{eqnarray}
In order to write the collinear gluon action, ultrasoft  gauge field $A^\mu_{us}$
is treated as a background field with respect to collinear gauge field $A^{n}_\mu$.
In this way, the QCD gluon action leads to the leading-order collinear gluon action in a covariant gauge as follows
\cite{scet}
\begin{eqnarray}
S_{ng}=\int d^4x\ \mbox{Tr}\left[\frac{1}{2g^2}([i\mathcal{D}^\mu, i\mathcal{D}_\mu] )^2   
+\tau ([i\mathcal{D}^\mu_{us}, A_{n{ }\mu}])^2+2\bar c_n[i\mathcal{D}^
\mu_{us}, [i\mathcal{D}_\mu, c_n] ]\right],
\end{eqnarray}
where $\tau$ is a gauge fixing
parameter for collinear gluon and   
\begin{eqnarray}
i\mathcal{D}^\mu &=&\frac{n^\mu }{2}(\bar{\mathcal{P}}+g\bar n\cdot A_n)+(\mathcal{P}^\mu_\perp +gA_{\perp,n}^\mu)+\frac{\bar n^\mu}{2}(in\cdot\partial+gn\cdot A_n+gn\cdot A_{us}),
\nonumber\\
i\mathcal{D}^\mu_{us}& =&\frac{n^\mu }{2} \bar{\mathcal{P}} + \mathcal{P}^\mu_\perp  +\frac{\bar n^\mu}{2}(in\cdot\partial +gn\cdot A_{us}).
\end{eqnarray} 

The lowest-order Faddeev-Popov action for  ultrasoft quarks and ultrasoft gluons  is   a  covariant gauge can be written by \cite{scet}
\begin{eqnarray}
S_{us}=\int d^4x  \left[\bar{\psi}_{us}i\slashed{D}_{us}\psi_{us}
-\mbox{Tr} \left(  \frac{1}{2} 
   G_{\mu\nu}^{us}G^{\mu\nu}_{us} +\tau_{us}  
( \partial_\mu A^\mu_{us})^2  +2  \bar c_{us} \partial_\mu  D^\mu_{us}
c_{us}  \right)\right],
\end{eqnarray}
where $\tau_{us}$ is a gauge fixing
parameter for ultrasoft gluon and $iD^\mu_{us}=i\partial^\mu+A^\mu_{us}$.

The complete Faddeev-Popov effective action for a single set of quark and gluon collinear modes in the $n$ direction, and quark and gluon ultrasoft modes in a covariant gauge
 is given by 
\begin{eqnarray}
S_{\mbox{scet}}=S_{n\xi}+S_{ng}+S_{us}.
\label{eff}
\end{eqnarray}
 We construct the following collinear and ultrasoft BRST transformations,  \\
(a) collinear BRST:
\begin{eqnarray}
\delta_b \xi_n &=& ic\xi_n\ \Lambda_n, \nonumber\\
\delta_b \bar\xi_n &=&- i\bar\xi_n c\ \Lambda_n, \nonumber\\
\delta_b A_n^\mu &=&[i\mathcal{D}^\mu, c_n]\ \Lambda_n, \nonumber\\
\delta_b c_n &=& \frac{g}{2}c_nc_n\ \Lambda_n,\nonumber\\
\delta_b \bar c_n &=&\tau  [i\mathcal{D}^\mu_{us}, A_{n{ }\mu}] \ \Lambda_n,
\label{brst1}
\end{eqnarray}
(b) ultrasoft BRST:
\begin{eqnarray}
\delta_b \psi_{us} &=& ic_{us}\psi_{us}\ \Lambda_{us}, \nonumber\\
\delta_b \bar\psi_{us} &=&- i\bar\psi_{us} c_{us}\ \Lambda_{us},\nonumber\\
\delta_b A_{us}^\mu &=& iD^\mu_{us}c_{us} \ \Lambda_{us},\nonumber\\
\delta_b c_{us} &=& \frac{g}{2}c_{us}c_{us}\ \Lambda_{us},\nonumber\\
\delta_b \bar c_{us} &=&\tau_{us}  i\partial^\mu  A^\mu_{us} \ \Lambda_{us}. 
\label{brst2}
\end{eqnarray}
under which the effective action in Eq. (\ref{eff}) remains invariant.
These sets of BRST symmetries  are very important in order to 
  renormalize the Feynman diagrams. With the help of these BRST transformations one can write 
  Slavnov-Taylor identities for partition function. 
The transformation parameters $\Lambda_n$ and $\Lambda_{us}$ are infinitesimal 
anticommuting parameters.
The generating functional for SCET action can be written by
\begin{eqnarray}
Z[0]=\int [D\phi_{n}][D\phi_{us}]\ \exp\left(iS_{\mbox{scet}}\right),\label{zen}
\end{eqnarray}
where $\phi_{n}$ and $\phi_{us}$ are generic notations for collective collinear and ultrasoft fields respectively. The collinear and ultrasoft path integral measures are invariant 
under corresponding BRST symmetry transformations.

\section{The FDBRST transformation}\label{3}

 \subsection{General setup}
 To construct FDBRST we define a generic notation for the   BRST transformations in Eqs. 
 (\ref{brst1}) and (\ref{brst2}) for a
collective field (having both collinear and ultrasoft fields) $\phi (x)$ as follows: 
\begin{equation}
\delta_b\phi (x) = s_{b}\phi (x)\ \Lambda,
\label{a}
\end{equation}%
where $s_{b}\phi $ is a Slavnov variation  and $\Lambda $ is an
global infinitesimal anticommuting parameter.  
 Following the standard procedure \cite{sdj}, a field-dependent
BRST transformation is constructed via interpolation of  a continuous parameter  $\kappa$ ($0\leq
\kappa \leq 1$) as: 
\begin{equation}
\frac{d\phi (x,\kappa )}{d\kappa}  =s_{b}\phi (x,\kappa )\Theta ^{\prime
}[\phi (\kappa )],  \label{b}
\end{equation}%
where  $\Theta^{\prime }[\phi (\kappa )]$ is an infinitesimal
field-dependent parameter.
In contrast to standard BRST transformation, this field-dependent transformation is not the symmetry of the path integral measure and amounts a precise Jacobian in the generating functional. This Jacobian contribution can be expressed as exponential of some functional of local fields and modifies the BRST exact part of the action \cite{sdj}.  The Jacobian of
functional measure is given by \cite{sud,sud1}
\begin{equation}
J[\phi ]={\ \exp \left( -\int d^{4}x\sum_{\phi }\pm s_{b}\phi (x)\frac{%
\delta \Theta ^{\prime }[\phi (x)]}{\delta \phi (x)}\right) }.  \label{J}
\end{equation}%
This Jacobian  therefore extrapolates the  action (within
functional integration) of the SCET theory   (\ref{zen}) as follows: 
\begin{eqnarray}
Z[0]\longrightarrow  \int [D\phi_{n}][D\phi_{us}]\ \exp\left(iS_{\mbox{scet}}- \int d^{4}x
 \sum_{\phi }\pm s_{b}\phi \frac{\delta
\Theta ^{\prime }[\phi ]}{\delta \phi }\right).
\end{eqnarray}%
This modified expression due to FDBRST  does not amount any changes in the physical content of the theory but rather
simplifies various issues in a dramatic way. In the next subsection we are going to demonstrate this.

\subsection{Collinear FDBRST transformation}

Following above methodology, we construct the   infinitesimal collinear FDBRST transformations as
\begin{eqnarray}
\frac{d \xi_n  }{d\kappa} &=& ic\xi_n\ \Theta ^{\prime }_n, \nonumber\\
 \frac{d \bar\xi_n  }{d\kappa} &=&- i\bar\xi_n c\ \Theta ^{\prime }_n, \nonumber\\
 \frac{d  A_n^{\mu} }{d\kappa} &=&[i\mathcal{D}^\mu, c_n]\ \Theta ^{\prime }_n, \nonumber\\
 \frac{d  c_n }{d\kappa} &=& \frac{g}{2}c_nc_n\ \Theta ^{\prime }_n,\nonumber\\
 \frac{d \bar c_n }{d\kappa} &=&\tau  [i\mathcal{D}^\mu_{us}, A_{n{ }\mu}] \ \Theta ^{\prime }_n,
\end{eqnarray}
where $\Theta ^{\prime }_n$ is an infinitesimal collinear field-dependent transformation 
parameter. This parameter can be chosen arbitrarily provided that must be nilpotent in nature. 
In the next section we will construct appropriate  $\Theta ^{\prime }_n$ to show how the  
generating functionals  corresponding to  various effective actions in different gauges are 
related.  
\subsection{Ultrasoft FDBRST transformation}
In the similar fashion the infinitesimal  ultrasoft FDBRST transformations are derived as
\begin{eqnarray}
 \frac{d  \psi_{us}}{d\kappa}  &=& ic_{us}\psi_{us}\ \Theta ^{\prime }_{us}, \nonumber\\
  \frac{d   \bar\psi_{us}}{d\kappa} &=&- i\bar\psi_{us} c_{us}\ \Theta ^{\prime }_{us},\nonumber\\
 \frac{d   A_{us}^{\mu}}{d\kappa}  &=& iD^\mu_{us}c_{us} \ \Theta ^{\prime }_{us},\nonumber\\
  \frac{d   c_{us}}{d\kappa} &=& \frac{g}{2}c_{us}c_{us}\ \Theta ^{\prime }_{us},\nonumber\\
 \frac{d   \bar c_{us} }{d\kappa} &=&\tau_{us}  i\partial^\mu  A^\mu_{us} \ \Theta ^{\prime }_{us},
\end{eqnarray}
where $\Theta ^{\prime }_{us}$ is an arbitrary infinitesimal ultrasoft field-dependent 
transformation parameter.

\subsection{Implementation of FDBRST transformation}

In this subsection, we assign  some specific values for the field dependent parameters $\Theta ^{\prime }_n$ and $\Theta ^{\prime }_{us}$  and calculate Jacobians of functional measures  under respective FDBRST transformations.
In this regard, we first choose the  parameter of collinear FDBRST transformation 
   $\Theta ^{\prime }_n$   as
\begin{eqnarray}
\Theta ^{\prime }_n =-i\int d^4y\  \mbox{Tr}\left( \bar c_n  [i\mathcal{D}^\mu_{us}, A_{n{ }\mu}]-\bar c_nf_1[A_{n{ }\mu}, A_{us}^ \mu] \right),
\end{eqnarray}
where $f_1[A_{n{ }\mu}, A_{us}^ \mu]$ is a most general collinear gauge condition.
For this parameter, the Jacobian of functional measure (\ref{J})  yields
\begin{equation}
J_n=e^{ -i\int d^{4}x\  \mbox{Tr} \left[  \tau ([i\mathcal{D}^\mu_{us}, A_{n{ }\mu}])^2+2\bar c_n[i\mathcal{D}^
\mu_{us}, [i\mathcal{D}_\mu, c_n] ] -f^2_1[A_{n{ }\mu}, A_{us}^ \mu] - 2\bar c_n [ \frac{df_1}{dA_{n{ }\mu}}, [i\mathcal{D}_\mu, c_n] ]\right]}.  
\end{equation}%
Here, we try to emphasize that we utilize an appropriate BRST transformation for the antighost
fields according to gauge conditions.
On the other hand the field dependent parameter  $\Theta ^{\prime }_{us}$  for ultrasoft BRST is chosen as
\begin{eqnarray}
\Theta ^{\prime }_{us} =-i\int d^4y\  \mbox{Tr}\left(\bar c_n  (i\partial_\mu A^\mu_{us})-\bar c_nf_2[ A_{us}^ \mu] \right),
\end{eqnarray}
The Jacobian of functional measure (\ref{J}) with this parameter leads to
\begin{equation}
J_{us}=e^{  i\int d^{4}x\  \mbox{Tr} \left[  \tau_{us}  
( \partial_\mu A^\mu_{us})^2  +2  \bar c_{us} \partial_\mu  D^\mu_{us}
c_{us} + f^2_2[  A_{us}^ \mu] +2\bar c_n   \frac{df_2}{dA_{us}^\mu} iD^\mu_{us}
c_{us} \right]}.   
\end{equation}%
These results eventually implies that under FDBRST transformation with the above field dependent parameters
\begin{eqnarray}
Z[0]=\int [D\phi_{n}][D\phi_{us}]\ \exp\left(iS_{\mbox{scet}}\right)\longrightarrow
\int [D\phi_{n}][D\phi_{us}]\ \exp\left(iS^f_{\mbox{scet}}\right), 
\end{eqnarray}
where final effective action is defined by
\begin{eqnarray}
S^f_{\mbox{scet}}=S_{n\xi}+S^f_{ng}+S^f_{us}. 
\end{eqnarray}
with  
\begin{eqnarray}
S^f_{ng}&=&\int d^4x\ \mbox{Tr}\left[\frac{1}{2g^2}([i\mathcal{D}^\mu, i\mathcal{D}_\mu] )^2   
+\tau f^2_1[A_{n{ }\mu}, A_{us}^ \mu]+2\bar c_n \left[ \frac{df_1}{dA_{n{ }\mu}}, [i\mathcal{D}_\mu, c_n] \right]\right],\\
S^f_{us}&=&\int d^4x  \left[\bar{\psi}_{us}i\slashed{D}_{us}\psi_{us}-
\mbox{Tr} \left( \frac{1}{2} G_{\mu\nu}^{us}G^{\mu\nu}_{us}-\tau_{us}   
f^2_2[  A_{us}^ \mu]-2 \bar c_n   \frac{df_2}{dA_{us}^\mu} iD^\mu_{us}
c_{us}\right)\right].
\end{eqnarray}
These are nothing but the leading-order collinear gluon action with gauge-fixing condition $f_1[A_{n{ }\mu}, A_{us}^ \mu]= 0$ and  leading-order action for  ultrasoft quarks and ultrasoft gluons with gauge-fixing condition $f_2[ A_{us}^ \mu]= 0$.
Thus,   FDBRST transformation upon implementation on generating functional 
changes the gauge-fixing  and ghost sectors of collinear and ultrasoft gluon actions.
This result will be very useful in handling the Feynman processes of the theory.
Since calculations of different Green's functions  depend on the choice of gauge-fixing condition and for some  particular choices the calculations are  simplified greatly, the structure of Faddeev-Popov action corresponding to that gauge can be achieved easily from this FDBRST formulation.

\section{Applications and conclusions} \label{4}

In this paper, we have considered an effective theory in light-cone coordinates which describes the interactions of soft and collinear modes in the presence of a hard interaction. 
By eliminating  the small components after decomposition of fermion field, we have written
 a gauge invariant SCET I action which admits different sets of
 gauge invariance in different momentum regions. In order to quantize correctly, we need to
 extend classical action   by adding suitable terms which break the local gauge invariance. 
 Such gauge variant terms attribute ghost terms in the generating functional of the theory.
 We have  developed two independent  sets of BRST symmetries which leave the Faddeev-Popov actions for collinear and ultrasoft sectors invariant separately. These BRST transformations 
 may help to write the counter terms to make the theory renormalizable. 
 
Furthermore, we
have   extended these sets of  BRST symmetries by making the transformation parameter
 field dependent. The difference of these extended symmetries to the usual one lies to the fact that these are not symmetry of the functional measure and, in contrast to the usual one,  eventually lead  to a local 
 Jacobian. On the physical ground, this Jacobian do not modify the theory as 
 all the changes attributed to the BRST-exact parts of the action. We have shown that
 for some specific choices of field-dependent parameters the exact expressions for
 various gauge-fixed actions   can suitably be  derived. These results are of particular importance  for the theoretical estimation of decay processes. This is because the 
 certain diagram calculations get simplified greatly in some particular gauge choices.
 For  instance,  it has been shown that by extending
  SCET formulation to the class of singular gauges, a
new Wilson line, the $T$  Wilson line, has to be invoked as a
basic SCET building block \cite{al,al1}. It is shown there that study in non-covariant gauges extend  the range of applicability of SCET. The transition from one gauge to another in SCET  can easily be done through our approach.

\end{document}